# Cardiac MRI Semantic Segmentation for Ventricles and Myocardium using Deep Learning

Racheal Mukisa and Arvind K. Bansal

Department of Computer Science
Kent State University, Kent, OH, USA
{rmukisa1, akbansal}@kent.edu

**Abstract.** Automated noninvasive cardiac diagnosis plays a critical role in the early detection of cardiac disorders and cost-effective clinical management. Automated diagnosis involves the automated segmentation and analysis of cardiac images. Precise delineation of cardiac substructures and extraction of their morphological attributes are essential for evaluating the cardiac function, and diagnosing cardiovascular disease such as cardiomyopathy, valvular diseases, abnormalities related to septum perforations, and blood-flow rate. Semantic segmentation labels the CMR image at the pixel-level, and localizes its subcomponents to facilitate the detection of abnormalities, including abnormalities in cardiac wall motion in an aging heart with muscle abnormalities, vascular abnormalities, and valvular abnormalities. In this paper, we describe a model to improve semantic segmentation of CMR images. The model extracts edge-attributes and context information during down-sampling of the U-Net and infuses this information during up-sampling to localize three major cardiac structures: left ventricle cavity (LV); right ventricle cavity (RV); and LV myocardium (LMyo). We present an algorithm and performance results. A comparison of our model with previous leading models, using similarity-metrics between actual image and segmented image, shows that our approach improves Dice similarity coefficient (DSC) by 2%-11% and lowers Hausdorff distance (HD) by 1.6 – 5.7 mm.

**Keywords:** Automated Segmentation, Cardiac MRI, Deep Learning, Image Processing, Medical Diagnosis, Semantic Segmentation, U-Net

## 1 Introduction

According to the statistics provided by the World Health Organization, cardiovascular diseases are the leading cause of death globally [1]-[3]; around 18 million deaths occur every year worldwide due to cardiovascular diseases [1]-[3].

The proper functioning of the heart is disrupted by cardiac abnormalities, such as irregular heart-beats, myocardial hypertrophy – thickening of the heart muscles, perforation in the septum dividing the left and right heart chambers causing deoxygenated blood to get mixed with oxygenated blood, valvular defects such as *stenosis* – abnormal narrowing negatively affecting blood-flow; *regurgitation* – blood flowing back through the valves due to improper synchronization or valve closing; and blood-leakage through perforated walls or holes in the septum [2][4]; congenital fusion of arteries with veins causing mix-up of oxygenated and deoxygenated blood; calcification in valves and arteries causing blood-flow blockage.

To accurately analyze cardiac abnormalities, there is a need for accurate delineation of the left ventricle (LV), right ventricular (RV), deformable wall motion analysis for end-diastolic (ED) and end-systolic (ES) phase, estimation of blood-flow rate from LV or RV, and synchronization analysis of ECG with valvular motion, and blood-ejection rate [4]-[6].

Echocardiography (EchoCG), Cardiac Magnetic Resonance (CMR) and Computed Tomography (CT) are non-invasive imaging techniques for diagnosing and localizing cardiac disease using image analysis [3][6]-[9]. The choice of imaging modality depends on multiple factors such as cardiac condition, patient characteristics, radiation exposure, cost, and the level of accessibility. EchoCG is the most widely used technique due to its availability and inexpensive nature [3][6]. However, EchoCG has lower image resolution compared to MRI and CT. CMR images have higher resolution modality [3][6]. CMR is regularly used for myocardial scar assessment, quantification of myocardial strain and intra-cardiac dynamics of blood-flow through the cardiac system [9].

For years, manual image segmentation has been the standard for left ventricle (LV), right ventricle (RV), and surrounding structures from CMR images [5][10][11]. Wall Motion Abnormality (WMA) analysis in the past has also been mainly performed by manually labelling the epicardium and endocardium contours to calculate the quantitative wall thickening [10][12]. However, manual segmentation is subject to inter-observer variation, variations arising due to irregular cardiac wall-motion, heart-rhythm and interpretation results based on radiologists' training and expertise [13].

The development of computationally efficient and accurate methods for automated cardiac segmentation, fibrous deformities, WMA analysis, and blood-flow analysis are much needed [12]. Cardiac image segmentation is an important first step that involves partitioning the image into semantically meaningful regions, based on which quantitative measures such as LV volume, RV volume, myocardial mass, wall thickness and ejection fraction (EF) can be extracted and analyzed [12]. Careful segmentation of the cardiac structures is essential for an accurate computation of cardiac parameters at the end of diastole (heart-relaxation) and systole (heart compression) [14].

The application of deep learning (DL) techniques for automated medical image segmentation is one of the recent advances aimed at assisting clinicians perform more accurate diagnoses and make informed medical decisions for medical interventions and treatments [13]. Several studies have exploited the use of DL techniques for automated cardiac segmentation in recent years [5][11][13]-[33]. However, limited research has been done on automated cardiac segmentation techniques that can label and accurately differentiate between different cardiac substructures such as LV, RV and myocardial region, and flexible Myocardial wall (Myo) and valvular motions. Researchers have mainly focused on segmentation of single substructures or deformity-based cardiac wall motions [7][11][13][15].

Automated detection of cardiac abnormalities requires semantic segmentation of different comprising substructures with different attributes and surroundings. For example, compared to LV-segmentation, the RV-segmentation is more complex due to RV-attributes such as its crescent shape, non-homogeneity in apical image slices, and relatively thinner wall [11]. Similarly, left ventricular WMAs are often visually assessed; their evaluation is based on systolic wall-thickening and systolic excursion criteria [4][9][10].

The major problems in earlier research studies can be categorized as: 1) deformation and contour-based segmentation start with some generic prior shape or initial contour that introduce unwanted bias; 2) substructure contours cannot be identified accurately by the proposed deep learning models due to varying features of substructures such as irregular shape, non-uniform brightness caused by blood-flow, different shape and motion of the substructures [11][14]. Deformity based techniques using prior shapes and initial contours suffer from built-in approximation in energy minimization. However, edge is a common feature that characterizes the defining contours of all major cardiac substructures. Preserving edge-information accurately will improve the accuracy of the substructure segmentation.

In this paper, we propose EE-UNet – a deep learning based automated semantic-segmentation model for labeling of three major cardiac substructures: the left ventricle (LV) cavity, the right ventricle (RV) cavity and the LV myocardium (Myo) from short-axis cardiac CMR images. Our approach infuses edges extracted during down-sampling phase of U-Net to the corresponding up-sampling layer to provide better context for the semantic segmentation and more accurate classification of different cardiac structures. Performance evaluation, as described in section 6, shows that the infusion of extracted edge-information significantly improves the substructure segmentation.

Major contributions of this research are:

1. Automated semantic segmentation of major cardiac substructures using U-Net;
2. Augmenting the U-net with edge-extraction to identify heart-substructures such as LV, RV and cardiac wall more accurately.

The rest of this paper is organized as follows. Section 2 describes the background concepts. Section 3 discusses work related to semantic segmentation. Section 4 describes the approach used in our work. Section 5 discusses work implementation details. Section 6 describes results and discussion. The last section concludes the discussion and outlines future works.

## 2    Background

### 2.1    Heart Anatomy

A heart is a muscle-complex that takes deoxygenated blood from the body, transports it to lungs for oxygenation, and pumps the oxygenated blood periodically around sixty times in a minute. It works based upon ion-variation (Ca++, Na+, K+) initiated voltage gradient that causes periodic muscle compression and relaxation [6].

The heart structure comprises four chambers: left atrium (LA); right atrium (RA); left ventricle (LV); right ventricle (RV). LA gets oxygenated blood from the lungs; LV pumps oxygenated blood to the body; RA collects deoxygenated blood from the body; RV sends deoxygenated blood to the lungs. The left and right chambers are separated by a muscle called 'septum'. The heart supplies blood through the main artery Aorta

[4][6]. The heart itself gets blood through two major arteries: left coronary artery (LCA) and right coronary artery (RCA) [4].

A heart has four valves: *aortic*; *mitral*; *pulmonary*; *tricuspid* [3][4]. *Aortic valve* is between left ventricle and the body and is involved in sending oxygenated blood to body; *mitral valve* is between left atrium and left ventricle and is involved in sending oxygenated blood obtained from the lungs to the left ventricle; *pulmonary valve* is between the right ventricle and the lung and is involved in transporting deoxygenated blood to the lungs; *tricuspid valve* is between right atrium and right ventricle and is involved in transporting deoxygenated blood from the body to the left ventricle. These valves work synchronously with the heart's compression and relaxation.

### 2.2    Cardiovascular Diseases

Cardiovascular diseases (CVD) are broadly categorized as: 1) thickening of the heart muscles due to calcification or old age which impacts blood-flow leading to stenosis (blood-leakage), regurgitation (backward blood-flow), and improper heart contraction and relaxation due to valvular deformities; 2) congenital heart diseases resulting in blockage of blood-vessels, perforation in the septum, or improper fusion of blood vessels, causing deoxygenation of blood in the body (ischemia) ultimately causing myocardial infarction (dead muscles due to cell-death); 3) rheumatic heart disease caused by permanent damage to the heart-valves from the bacterial infection; 4) thrombosis and embolism in the heart resulting in valvular lesions in the aorta-valve and weakening of the valve-function [3][4]. A hole in the septum contaminates oxygenated blood in the LV (or LA) with deoxygenated blood in the RV (or RA).

Proper identification of cardiac abnormalities requires CMR analysis of the heart muscles, cardiac wall-motion analysis, including valvular motion during heart contraction and relaxation, estimation of blood-flow analysis through arteries and various heart chambers, and arterial blockages.

### 2.3    Semantic Segmentation

Classical image segmentation derives a set of homogeneous regions in an image with no labeling. Semantic segmentation maps pixels in an image to the corresponding labels without differentiating object instances [24][29][34]; it classifies similar objects as a single class. Semantic segmentation facilitates image classification, object detection, and boundary localization. Semantic features are essential in dense recognition tasks such as semantic segmentation [34].

### 2.4    Deep Learning Techniques Related to Cardiac Segmentation

Deep learning (DL) is a class of recent (last two decades) automated learning techniques, not covered under classical machine learning, for efficient natural language translation, image classification, and object detection using a combination of one or more techniques such as local and global feature analysis, context identification by providing feedback from the previous output(s) of neural network, and probabilistic prediction of the next element of the occurrence in a sequence using previous subsequence and the probability obtained from statistical analysis of a large database of similar sequences along with the use of deep feed-forward neural network with multiple hidden layers or fully connected network [24][35]-[40].

Popular subclasses of DL techniques related to semantic segmentation are: 1) Convolutional Neural Network (CNN) and variants such as U-Net [17]; 2) Recurrent Neural Network (RNN) – provides feedback from past output along with input to provide short-term context [35]; 3) Long Short-Term Memory – a variant of RNN to avoid error magnification or context loss in RNN by using a forget-gate [35]; 4) self-attention based auto-encoders – a probabilistic model that encodes the input signals to generate tokens and decodes the sequence of token, previously derived probability and feed-forward neural network with a large number of hidden layers [35]; 5) Fully Convolutional Network (FCN) – a statistical learning technique based on auto-encoders [24]; 6) Transformers – a probabilistic learning technique that uses a large database to generate probability of the occurrence of the next entity in the sequence of entities.

CNN comprises a stack of feature-map extractors followed by an FNN or FCN [6]. Feature-map extractor comprises a stack of three types of layers: *convolution filters* to extract local features, *RELUs* to denoise; *pooling unit* for summarization and dimension reduction [6]. The feature-vector produced by the stack goes through FNN or FCN for classification [6]. The major application of CNN is in image classification, and identification. In CMR analysis, CNN and variants like U-Net are used for cardiac segmentation to estimate the object of interest and for semantic segmentation [24].

Fully Convolutional Networks (FCN), is a variant of CNN that uses an inbuilt encoder-decoder network for mapping input to high-level features and extracting information to make predictions [24]. FCNs are trained and applied to the entire image dataset, and output images of the same size as input.

The U-Net is a semantic segmentation network based on FCN [24]. With an inbuilt contracting and expansive path, the encoder performs all high-level feature-extraction tasks through down-sampling and convolutional layers, and the decoder transforms those high-level features into semantic labels through up-sampling. U-Net employs skip-connections between the corresponding steps in down-sampling and up-sampling to recover spatial context lost during the down-sampling path. Feature maps from the encoding paths are concatenated to the decoding paths to add fine-grained features into the dense predictions. Compared with FCN, U-Net is computationally more efficient because there is no fully connected layer in the structure. The U-Net has found major use in biomedical image analysis [19][21][26].

## 3   Related work

Multiple deep learning techniques such as CNN, FCN, U-Net, auto-encoders, vision transformers, their variants, and combinations are being experimented for accurate cardiac segmentation [2][11]-[16][19]-[29][31]-[34]. We classify related work for cardiac image-segmentation as: 1) deformable and active contour models [7][20]; 2) CNN variants based semantic segmentation [11][14][33]; 4) FCN based segmentation [11][14][23][28][31]; 5) UNet based semantic segmentation [19][21][26][40]; 6) Vision transformer based semantic segmentation [29][39]. In addition, there is limited research based upon sparse matrix analysis of cardiac wall motion by Fung et al. to analyze wall-motion abnormalities [30].

Podlesnikar et al. [7] study the strain of the major features by tracking the movement of major features in radial, circumferential and longitudinal directions. Such a feature-tracking would require automated segmentation and analysis of features such as LV, RV and Myo. However, they do not have a clear technique for automated segmentation. Deformable models such as active contours and level-set methods evolve by using the energy function to minimize their energy during segmentation. The energy minimization model uses approximation and lacks accuracy.

Avendi et al. [11] proposed the integration of CNN and auto-encoders for automated segmentation of right ventricle (RV) in cardiac short-axis MRI. Their model integrated with deformable models converges faster compared to conventional deformable models. However, they are unable to capture the contours of LV, RV, Myo accurately as they do not extract edge-attributes accurately. Besides, as pointed out by the researchers RV segmentation suffers from the irregular shape and surroundings [11][14].

Zotti et al. [33] combine the knowledge of prior shape of the substructures and CNN to label the structures. They report accuracy using Dice coefficient of .91 and Hausdorff distance as 9.5 mm. Our performance result shows much better accuracy because we use the actual contour instead of a generic shape as explained in subsection 6.2.

Long et al. [24] used the FCN for semantic image-segmentation by adapting the contemporary classification networks that were fine-tuned for the segmentation task. A major challenge with FCN-based networks is that the resolution of the input image reduces by a large factor during down-sampling making it difficult to reproduce the finer image details after up-sampling. This lack of finer details results in the loss of the substructure contours. Our scheme extracts the edge information in the beginning and infuses the edge-information during U-Net expansion phase to preserve the contour information of the substructure.

Khened et al. [23] propose an efficient version of FCN that optimizes the skip connections during up-sampling. They extract handcraft-features used by the medical practitioners during diagnosis and use spatio-temporal statistics of cardiac structures and Hough transforms to train for Region of Interest (ROI). They also label three major substructures: LV, RV and Myo. Compared to their approach, we use only edge-attributes extraction for the segmentation. The comparison with their approach shows that our scheme performs much better both in terms of Dice Similarity Coefficient (DSC) and Hausdorff distance (HD (see Table 3, Section 6.2). As described earlier, the use of FCN in their work loses finer image details by a large factor during down-sampling.

Grinias et al. [22] use an active contour model between two stable phases end-diastolic (ED) and end-systolic (ES) and use the energy minimization function to approximate the contour. Their scheme also suffers from energy minimization approximation. In contrast, our scheme is based upon extracting edge-attributes from the images. The comparison shows that our scheme outperforms their scheme (see Table 3, Section 6.2).

Our research experimented with U-Net augmented with edge-extraction (see section 4.2), dense nets and attention network models to perform the semantic cardiac segmentation task. As described in section 6, EE-UNet architecture proposed in this research achieved better results and stability than dense nets and attention networks.

## 4    Edge Enhanced UNET (EE-UNet) Model Architecture

For cardiac image segmentation, the typical anatomic structures of interest often include the LV, RV, LA, RA, and heart vessels. In Figure 1, we categorize different cardiac image segmentation tasks performed based on an imaging modality used for capturing cardiac images.

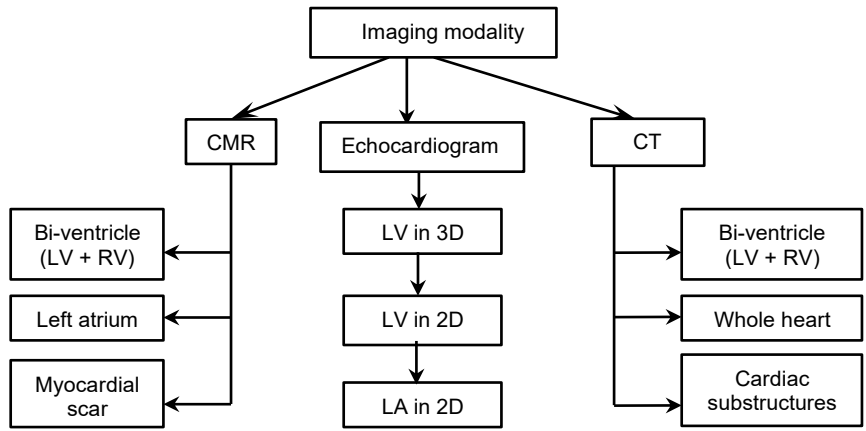

**Fig 1.** Cardiac image segmentation tasks for different imaging modalities

We develop a DL-based semantic cardiac CMR segmentation based on the integration of U-Net architecture with the edge information extracted during the down-sampling phase and infused in the up-sampling phase. The infused extracted-edge information helps to accurately localize boundaries between different segmentation classes such as LV, RV, LV-Myocardium (Myo), and the background, given that all the classes exist next to each other. This in turn enhances our model's accuracy of segmenting cardiac ventricles and the myocardium from CMR.

Our method is implemented in three phases, as illustrated in Figure 2. The steps are: 1) preprocessing; 2) down-sampling and edge-extraction during contraction phase; 3) infusing the extracted edge and context information during the expansion phase.

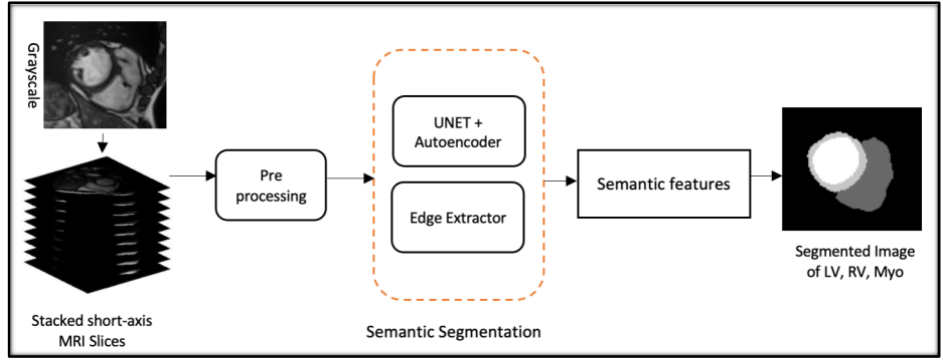

**Fig. 2**. EE-UNet based cardiac semantic segmentation model

### 4.1 EE-UNet Model

Automatic cardiac segmentation is implemented in our model to extract four semantic classes from each image: the background; the LV cavity; RV cavity; the LV myocardium. Figure 3 illustrates our proposed EE-UNet model.

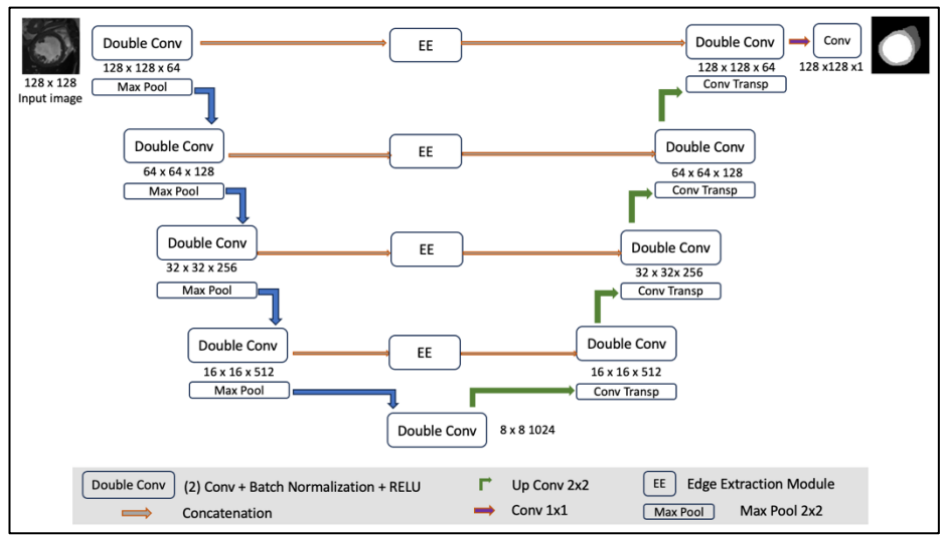

**Fig 3**. The proposed EE-UNet architecture

EE-UNet has a built-in edge extraction module that extracts edge-features and context information in the spatial domain during the model's contracting path. This information is again integrated with the corresponding steps in expansion path.

On the left side of the model is the contacting path and on the right side is the expansive path. Given *N* number of varying image slices that can be passed as input channels to the model, every step in the model's contracting path comprises four convolution blocks of 64, 128, 256, and 512 kernel sizes.

Each convolution block has two successive 3x3 convolution layers followed by their respective batch normalization layers and the *rectified linear unit* (*ReLU*) activation function. The contracting path down-samples the image by reducing the spatial dimensions at every layer while doubling the channels with a 2x2 max pooling operation of stride 2. A skip connection is then stacked to the second convolution block next to the first convolution block. An edge extraction module is infused at each layer as detailed in section 4.3. The bottleneck layer connects the contracting path and the expansive path, and it comprises one convolutional block of 1024 kernel size.

The expansive path comprises four up-convolution blocks of 512, 256, 128 and 64 kernels, and it increases the image spatial dimensions while reducing the channels. Each up-convolution block comprises two successive 2x2 convolution transpose-layer whose feature-map is concatenated with the output from the edge extractor from the corresponding layer in the contracting path. The final convolution layer outputs the segmented masks corresponding to the input images.

Using the dice loss-function, the output from the last layer is used to compute the loss. Once the loss is computed, ADAM optimizer is used to optimize it. Then, the model parameters are updated through backward propagation.

### 4.2   Edge-extraction Module

The edge extraction algorithm first denoises and smoothens the image using a Gaussian function (a weighted average of neighboring pixels). Then, it computes the edge-length $\lambda_{it}$ using eqns. (1)-(3) where $i$ and $j$ represent locations for each pixel, and $I$ is the intensity value of the pixel. Edge-gradient is computed by eqn. (4).

$$\delta_{ij}^{x} = \frac{I_{i(j+1)} - I_{ij} + I_{(i+1)(j+1)} - I_{(i+1)j}}{2} \tag{1}$$

$$\delta_{ij}^{y} = \frac{I_{ij} - I_{(i+1)j} + I_{i(j+1)} - I_{(i+1)(j+1)}}{2} \tag{2}$$

$$\lambda_{ij} = \sqrt{(\delta_{ij}^{x})^2 + (\delta_{ij}^{y})^2} \tag{3}$$

$$\theta_{ij} = tan^{-1}\left(\frac{\delta_{ij}^{y}}{\delta_{ij}^{x}}\right) \tag{4}$$

The edge extraction algorithm uses the canny-edge method, adopted to refine the image-edges by computing the position of the edge for the target class using the image-gradient. The canny operator is more robust compared with other operators and yields thin edges making it easy to locate edge-positions and improve the signal-to-noise ratio.

### 4.3 Algorithm

The overall algorithm for the semantic segmentation is given in Figures 4. The subscripts i and j show x and y coordinate of the pixel. The semantic segmentation algorithm accepts a dataset of batches of (*image, mask*) pairs. For each epoch *e*, the initial model ω is continuously updated after each image is analyzed using backpropagation and the loss value $1 - Dice^{weighted}$ (see subsection 6.1) after recognizing the segmented substructure and the corresponding label for every image. After all the epochs are over, the updated model ω is returned.

---

**Algorithm** Semantic segmentation
**Input: 1.** Augmented dataset *d* : a sequence of batches.
        Each batch is a sequence of (image *img*, mask *m*) pairs
   2. initial model ω, 3. learning rate ρ; 4.batch bs;
   5. no_of_epochs *e*; 6. kernel_size k
**Output:** trained_model ω;
**{for each** epoch *i* (1 ≤ i ≤ e) {
   **for each** batch ∈ d {
      **for** each pair*(img, m)* ∈ batch **{ %** process one image at a time
         featureSet = img;
         **for each** downlayer in **encoder** until the bottom_layer { % down-sample
            feature$^{down}$ = down_sample(*featureSet*, k);
            E = **edge_extract**(feature$^{down}$); edgeSet S$^{edge}$ = E + S$^{edge}$;
            feature$^{pooled}$ = max_pooling (feature$^{down}$);
            feature$^{Set}$ = feature$^{pooled}$;}
         Let the bottom feature-set be feature$^{bottom}$; feature$^{recovered}$ = feature$^{bottom}$;
         **for each** uplayer in decoder until finished {
            E = first(S$^{edge}$); feature$^{recovered}$ = upsample (feature$^{recovered}$)
            feature$^{recovered}$ = infuse(feature$^{recovered}$, E);
            S$^{edge}$ = rest(S$^{edge}$);}
         Let the predicted image be img$^{pred}$;
         Calculate DSC using the eqn. (5) in section 6
         loss = 1 – DSC; % optimize Loss using Adam optimizer;
         ω = backpropagation( ω, ρ, loss) % update the model using backpropagation
      }}}
   **return** the final model ω;
}

---

**Fig. 4.** An algorithm for semantic segmentation using EE-UNet

Each pair (image *img*, mask *m*) is down-sampled iteratively to extract the feature-maps *feature*$^{down}$ which are processed by a pooling unit to generate concentrated feature-set feature-map *feature*$^{pooled}$.

Before applying the pooling layer, edge-label and edge-attributes are extracted by applying canny operator to *feature*$^{down}$. This edge information is stored in a stack $S^{edge}$ iteratively until the bottom-layer of the U-Net is reached.

After the bottom-layer is reached, up-sampling is used to reconstruct the segmented image layer by layer iteratively. At each layer, edge-information from the stack $S^{edge}$ is popped, and infused in the recovered feature-map *feature*$^{recovered}$. The process is repeated until the stack $S^{edge}$ is empty. After the segmented image is reconstructed, loss is computed, and the current model ω is updated.

Edge-extraction algorithm (see Fig. 5) accepts as input a pair (image *img*, mask *m*) and outputs an edge-tuple of three attributes $E$ = (**label** *edge_pred*, **edge-length** $\lambda_{ij}$, **edge-orientation** $\theta_{ij}$). The label edge_pred could be *LV_edge, RV_edge,* or *Myo_edge*. For each pair of the form (*img, m*), the parameters *kernel_filter_size r* and *Gaussian_spatial_weight w* are derived. Denoising is done using Gaussian filters applied on the image *img* with the parameters r and w. The resulting denoised image *img*$^{denoised}$ is used for edge-extraction using canny operators.

```
Algorithm edge_extract
Input: (image img, mask m) pair
Output: 1. edge E = (label edge_pred ∈ {LV_edge, RV_edge, Myo_edge}, edge-length
                    λij, edge-orientation θij)
{ Filter_size r = scale the kernel size to match the image size;
  gaussian_spatial_weight w = (a constant × r) / 2     % constant is empirically decided
  denoised image Imgdenoised = apply_gaussian_filter(Img, r, w);
  edge E = apply_canny_operators(image Imgdenoised);
  ( δxij , δyij ) = apply_edge_thinner ( δxij , δyij );  %apply edge thinning
  calculate the length λij and the slope θij of the edge-segment E using eqn. 3 and 4;
  E = (edge_pred, λij, θij); return E;
}
```

**Fig. 5**. An algorithm for edge extraction

The extracted edge $E$ is further thinned using edge-thinners for better accuracy in deriving edge-length $\lambda_{ij}$ and the edge- orientation $\theta_{ij}$ using eqn. (3) and (4), respectively. Indices *i, j* denote the pixel-locations of the edge-starts. The resulting edge information are returned to the semantic segmentation algorithm.

## 5. Implementation

The proposed model was implemented and the experiments were performed on a Dell Server running the ubuntu 22.4 with a GPU of GEForce RTX 2070 using PyTorch, a popular DL framework written in python.

The dataset used to evaluate our proposed method is publicly available from the MICCAI 2017 automated cardiac diagnostic challenge (ACDC) database [38]. The dataset also consists of short-axis cine MR images of 150 patient exams acquired on 1.5-Tesla and 3-Tesla systems. The dataset includes manual expert segmentation of the RV and LV cavities, and LV myocardium to provide accurate ground-truth information.

The cine MR images were acquired in breath-hold mode with a prospective gating and with an SSFP sequence in short-axis orientation. All the patients are clinically diagnosed into five classes (4 pathological and 1 healthy subject classes) namely: 1) normal (NOR); 2) patients with previous myocardial infarction (MINF); 3) patients with dilated cardiomyopathy (DCM); 4) patients with hypertrophic cardiomyopathy (HCM); 5) patients with abnormal right ventricle (ARV).

Thirty cases are provided in each class. Each class is also defined based on physiological parameters such as the left or right diastolic volume or ejection fraction (EF), the local contraction of the LV, the LV mass, and the maximum thickness of the myocardium. For each patient, the weight, height, and the diastolic and systolic phase instants are provided.

### 5.1    Data Pre-processing

All the short-axis cine MR images are originally in 3D NIFT format with 12-35 frames available for each patient. Each ED and ES frames was clearly labeled. The original data contained some severe slice misalignments that originate from the different breath hold positions that exist when the slice stacks are being captured.

Due to these data inconsistencies, data preprocessing is an important step to ensure that our model is fed with similar input. A series of short-axis slices cover the LV from the base to the apex, with a slice thickness of 5 to 8 mm and an inter-slice gap of 5 or 10 mm. Also, short-axis MR images comprise a stack of 2D MR images acquired over multiple cardiac cycles which are often not perfectly aligned. Figure 6 shows sample ED and ES short-axis CMR image-frames from the dataset.

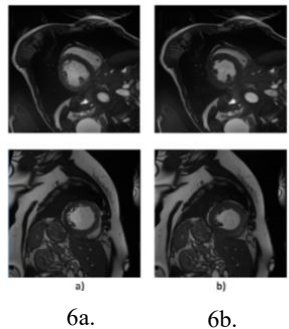

6a.            6b.

**Fig 6**. Sample short-axis cardiac CMR images: (6a) end-diastole frame; (6b) end-systole frame.

To correct for differences in gray-scale voxel size, all 3D slices were converted to 2D format by taking all the slices for both the original image and the mask as input. The resulting image-slice was pre-processed to output the corresponding 2D image in gray-scale, scaled to 128 x 128.

### 5.2    Training

The input images and their corresponding masks were used as input to train our model. To train, we used the 5-fold cross validation (4-folds for training, 1-fold for testing) repeated five-times. The model was trained using the Adaptive Momentum Estimation (ADAM) optimizer with a learning rate of 0.001 and other default parameters. We used the dice loss-function with a minimum batch gradient and batch-size of 32 to compute the loss. The training was completed after 100 epochs and the best models were archived for testing.

## 6. Results and Discussion

For cardiac diagnosis, digital images are the basic tool used for the computation of subsequent clinical indices from the shape and Haar-structure. We evaluated the performance of the proposed EE-UNet model using distance and clinical metrics.

The segmentation performance of the proposed method is evaluated using the mean *Dice Similarity Coefficient* (DSC) and *Hausdorff Distance* (HD) which provides information of similarity between obtained segmentations for LV, RV, and Myo with the reference ground-truth masks. Let X and Y be the segmentation result and ground truth, respectively [42]. The *DSC(X, Y)* is defined in eqn. (5).

$$DSC = 2 \times \frac{|X \cap Y|}{|X| + |Y|} \quad (5)$$

The parameter-value *DSC = 0* represents the zero overlap between the ground-truth and the derived segmentation; *DSC* equal to *1* denotes complete overlap between ground-truth and segmentation result in both the foreground and background.

*HD(X, Y)* is defined as the greatest of all distances from a point in one set to the closest point in the other set. It is defined in eqn. (6).

$$HD(X,Y) = \max_{x \in X} \min_{y \in Y} \| x\text{-}y \|_2 \quad (6)$$

where X and Y are two contour-point sets. A lower value of HD(X, Y) indicates that the distance between the ground-truth and the segmentation result is less and the two images are more similar. HD(X, Y) = 0 means that the segmentation result is accurate, as the boundaries of the segmented region closely resembles the true boundaries.

### 6.1 Loss Function

The Dice's Coefficient is a metric used to measure the similarity between two given samples. In this work, we calculated the weighted dice loss to deal with the complex nature of segmenting CMR images and used weights to offset the class imbalance. The loss function is calculated using eqn. (7).

$$Dice^{weighted} = 1 - \frac{2 \times \sum_p w^p D^p}{\sum_p w^p} \quad (7)$$

where $w^p$ is the weight assigned to class *p* and $D^p$ is the dice-score for class *p*. The dice score $D_p$ is computed as shown in eqn. (8).

$$D^p = 1 - \frac{2 \times \sum x_i y_i}{\sum x_i + \sum y_i} \tag{8}$$

where $p_i$ is the predicted output of the model and $x_i$ is the ground truth. Loss is calculated as 1 - $Dice^{weighted}$ and is used to train the model. Higher dice coefficient shows better prediction accuracy of the model.

### 6.2  Performance Evaluation

Table 1 shows Dice coefficients and Hausdorff distance results of 5-fold cross validation for model segmentation of LV, RV and LV myocardium at end-diastole and end-systole phases based on five pathological classes. (DCM = patients with dilated cardiomyopathy, HCM = patients with hypertrophic cardiomyopathy, MINF = patients with previous myocardial infarction, NOR = normal patients, ARV = patients with abnormal right ventricle).

In comparison to the reference standards, the left ventricle performed the highest at ED, and the right ventricle performed the lowest at ES. The results from Tables 1 and 2 show that the performance was substantially higher in ED than in ES for LV and RV. However, the performance was lower for Myo in ED and ES compared to LV and RV. The average Hausdorff distances were lower for LV than for RV. This could be due to the strong contrast between LV and the surrounding. In addition, the shape of RV is more irregular, which may contribute to this result.

**Table 1.** Cross-validation for EE-UNet model segmentation of five pathological classes. The values are rounded to the nearest second decimal place.

| Class | Instance | Dice score | | | Hausdorff distance(mm) | | |
|---|---|---|---|---|---|---|---|
| | | LV | RV | MYO | LV | RV | MYO |
| **DCM** | ED | 0.96 | 0.88 | 0.81 | 3.21 | 27.12 | 6.30 |
| | ES | 0.95 | 0.83 | 0.83 | 4.03 | 25.08 | 7.88 |
| | **Average** | **0.96** | **0.85** | **0.82** | **3.62** | **26.1** | **7.09** |
| **HCM** | ED | 0.90 | 0.81 | 0.84 | 5.04 | 26.22 | 6.93 |
| | ES | 0.81 | 0.75 | 0.87 | 7.43 | 24.79 | 5.06 |
| | **Average** | **0.85** | **0.78** | **0.85** | **6.23** | **25.50** | **6.00** |
| **MINF** | ED | 0.95 | 0.81 | 0.80 | 6.50 | 23.23 | 9.03 |
| | ES | 0.93 | 0.72 | 0.79 | 7.00 | 25.85 | 13.25 |
| | **Avg** | **0.94** | **0.76** | **0.80** | **6.75** | **24.54** | **11.14** |
| **NOR** | ED | 0.94 | 0.89 | 0.80 | 6.20 | 16.77 | 8.32 |
| | ES | 0.92 | 0.84 | 0.88 | 7.85 | 19.40 | 6.46 |
| | **Avg** | **0.93** | **0.86** | **0.84** | **7.02** | **18.08** | **7.39** |
| **ARV** | ED | 0.95 | 0.92 | 0.79 | 2.01 | 12.26 | 10.95 |
| | ES | 0.89 | 0.88 | 0.83 | 5.89 | 16.70 | 6.33 |
| | **Avg** | **0.92** | **0.90** | **0.81** | **3.95** | **14.48** | **8.64** |

Tables 2 shows segmentation results of LV, RV and LV myocardium at end-diastole and end-systole phases based on the entire dataset. Results show that performance was better in ED than in ES. Average Hausdorff distances were significantly lower for the LV than for the RV and myocardium. This might be caused by the strong contrast between the LV and the surrounding myocardium.

**Table 2.** Segmentation accuracy at end-systole (ES) for the LV, RV and Myo

|  | LV | | RV | | Myocardium | |
| --- | --- | --- | --- | --- | --- | --- |
|  | Dsc | Hd | Dsc | Hd | Dsc | Hd |
| **End-diastole** | 97.13 | 5.96 | 96.00 | 10.01 | 91.25 | 8.50 |
| **End-systole** | 94.30 | 6.70 | 92.95 | 9.70 | 89.76 | 9.60 |

We compared our results with the semantic segmentation results of four other studies by Grinias et al. [22], Khened et al. [23], Painchaud et al. [25], and Zotti et al. [33]. Table 3 summarizes the comparison results. The results show that DSC scores and HD scores in our EE-UNet model are better than all the leading models for both end-systole and end-diastole semantic segmentation. The only exception is Myo end-systole DSC scores of Zotti et al. and end-diastole data of Painchaid et al.

**Table 3.** Comparing the state-of-the-art performance on ACDC dataset. The result has been rounded to the first place after the decimal. ED denotes end-diastole; ES denotes end-systole.

| Methods | LV | | | | RV | | | | MYO | | | |
| --- | --- | --- | --- | --- | --- | --- | --- | --- | --- | --- | --- | --- |
|  | DSC | | HD | | DC | | HD | | DSC | | HD | |
|  | ED | ES | ED | ES | ED | ES | ED | ES | ED | ES | ED | ES |
| Khened | 96.4 | 91.7 | 8.1 | 9.0 | 93.5 | 87.9 | 14.0 | 13.9 | 88.9 | 89.8 | 9.8 | 12.6 |
| Zotti | 95.7 | 90.5 | 6.6 | 8.7 | 94.1 | 88.2 | 10.3 | 14.0 | 89.0 | 90.0 | 9.6 | 9.3 |
| Grinias | 95.0 | 85.0 | 9.0 | 13.0 | 89.7 | 77.0 | 19.0 | 24.2 | 80.0 | 78.4 | 12.3 | 14.6 |
| Painchaud | 96.1 | 91.0 | 6.1 | 8.3 | 93.0 | 88.0 | 13.7 | 13.3 | 88.4 | 90.0 | 8.6 | 9.6 |
| **EE-UNet** | **97.1** | **94.3** | **6.0** | **6.7** | **96.0** | **93.0** | **10.0** | **9.7** | **91.2** | 89.6 | **8.5** | 9.6 |

Figure 7 shows sample results from segmenting the three key cardiac structures: the LV, RV and Myo separately. Figure 8 lists visual samples of segmentation predictions obtained when all the three structures are combined.

LV shape in the short-axis CMRs is more clearly defined as compared to RV whose attributes have crescent shape and relatively thinner wall. Hence, segmentation of RV is more challenging for the algorithm to learn. The segmentation of myocardium is more challenging than the segmentation of LV and RV cavities, as reflected by lower *DSC*s for both *end-diastole* and *end-systole* (see Table 2).

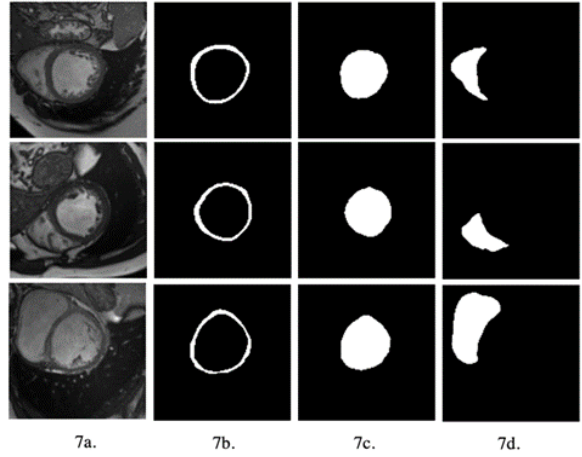

**Fig. 7**. The first column shows the original structure images. The subsequent columns show the corresponding segmented structures derived by our method: (a) original image (b) segmented Myo (c) segmented LV (d) segmented RV.

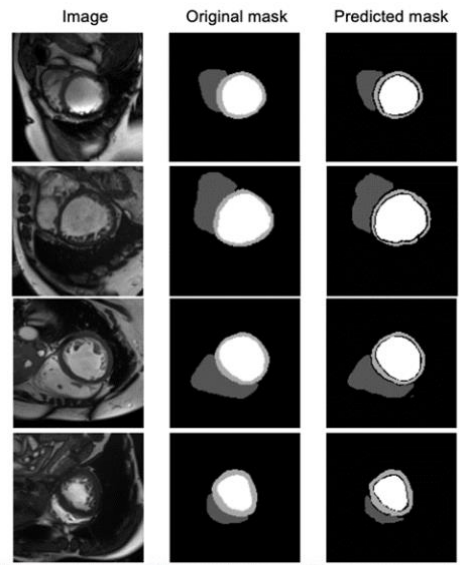

**Fig 8.** Visual examples of segmentation predictions from the proposed model

## Conclusion and Future Work

In this paper, we propose a modified 2D U-Net-based network infused with extracted edge-information for the semantic segmentation of cardiac ventricles (LV, RV) and myocardial scar (LMyo) from short-axis cine CMR. Edge-infusion ensures accurate context information and the localization of boundaries between different semantic classes, improving segmentation accuracy – DSC by 2%-11% and HD by 1.6 – 5.7 mm. The described approach is accurate and robust compared to other state-of-the-art methods. The performance of our model shows potential for its viable application in automated analysis to detect cardiac diseases.

We are extending this work to 3D segmentation, flexible heart-muscle analysis, and valvular motion analysis using synchronization with PQRST waveforms in ECG (Electrocardiograph).